\documentclass[hidelinks,onefignum,onetabnum]{siamart250211}
\usepackage{lipsum}
\usepackage{amsfonts}
\usepackage{graphicx}
\usepackage{epstopdf}
\usepackage{algorithmic}
\ifpdf
  \DeclareGraphicsExtensions{.eps,.pdf,.png,.jpg}
\else
  \DeclareGraphicsExtensions{.eps}
\fi

\newsiamremark{remark}{Remark}
\newsiamremark{hypothesis}{Hypothesis}
\crefname{hypothesis}{Hypothesis}{Hypotheses}
\newsiamthm{claim}{Claim}
\newsiamremark{fact}{Fact}
\crefname{fact}{Fact}{Facts}

\headers{Computational Fluctuating Hydrodynamics}{A.L. Garcia, J.B. Bell, A. Nonaka, I. Srivastava, D. Ladiges, AND C. Kim}

\title{An Introduction to Computational Fluctuating Hydrodynamics\thanks{Submitted to the editors DATE.
\funding{This work was funded by U.S.\ Department of Energy, Office of Science, Office of Advanced Scientific Computing Research, Applied Mathematics Program under contract No.\ DE-AC02-05CH11231.}}}

\author{Alejandro L. Garcia\thanks{San Jose State University, San Jose, CA (\email{Alejandro.Garcia@sjsu.edu})}
\and John B. Bell\thanks{Lawrence Berkeley National Lab, Berkeley, CA}
\and Andrew Nonaka\footnotemark[3]
\and Ishan Srivastava\footnotemark[3]
\and Daniel Ladiges\footnotemark[3]
\and Changho Kim\thanks{University of California Merced, Merced, CA}
}

\usepackage{amsopn}

\ifpdf
\hypersetup{
  pdftitle={An Example Article},
  pdfauthor={D. Doe, P. T. Frank, and J. E. Smith}
}
\fi

\usepackage{soul} % Use \st{} for strikeout, \ul{} for underline

\newcommand{\commentout}[1]{}

\newcommand{\half}{{\frac12}}
\begin{document}

\maketitle

% REQUIRED
\begin{abstract}
This short article is an introduction to the theory of fluctuating hydrodynamics (FHD) and the formulation of numerical schemes for the resulting stochastic partial differential equations.
The stochastic heat equation (SHE) is derived, and simple finite-volume schemes for its solution are outlined.
Numerical results from a Python program (see Supplementary Materials) are shown to be in good agreement with theoretical solutions of the SHE for equilibrium and non-equilibrium systems.
Finally, a menagerie of other FHD equations is outlined, and suggested instructional exercises are listed.
% This is an example SIAM \LaTeX\ article. This can be used as a
% template for new articles.  Abstracts must be able to stand alone
% and so cannot contain citations to the paper's references,
% equations, etc.  An abstract must consist of a single paragraph and
% be concise. Because of online formatting, abstracts must appear as
% plain as possible. Any equations should be inline.
\end{abstract}

% REQUIRED

% REQUIRED

\section{Introduction}

Partial differential equations (PDEs) are used to model the dynamics of gases, liquids, and solids.
Applications are found in biology, chemistry, physics, and every branch of engineering, and for all but the simplest PDEs we rely on numerical methods to solve these equations.
For example, numerical solution of the Navier--Stokes equations of hydrodynamics is used for general circulation models that simulate global weather, design of aircraft, and modeling combustion processes in engines, to name a few.
Due to the importance of numerical methods for solving PDEs, it is a standard topic in the undergraduate and graduate curricula for applied mathematics \cite{PDEtextbook2025}.

The diffusion equation is a simple example of a parabolic PDE that can be solved by a variety of analytic (e.g., separation of variables) and numerical (e.g., finite volume) methods. 
It models the transport of molecules by diffusion or that of energy by thermal conduction.
For the latter, one assumes that the heat flux is negatively proportional to the gradient of temperature.
This relation is known as the Fourier law, which is simply Newton's law of cooling in the continuum limit.
Although this simple model is accurate at the macroscopic scale, heat flow is more complicated at the microscopic scale.
More specifically, due to consistent random motion of molecules constituting the material, small variations of temperature are observed even when the system is at equilibrium.

Fluctuating hydrodynamics (FHD) was originally introduced by Landau and Lifshitz~\cite{Landau_59} as a way to put the aforementioned thermal fluctuations into a continuum framework by including a stochastic forcing to each dissipative transport process (e.g., the heat flux). 
Although FHD has proven to be useful in modeling transport and fluid dynamics at the mesoscopic scale, theoretical calculations have been feasible only with simplifying assumptions~\cite{Zarate_07}.  % \IS{cite Zarate \& Sengers book?}. 
As such, there is great interest in numerical schemes for computational fluctuating hydrodynamics (CFHD).
There are a variety of algorithms (e.g., finite element~\cite{martinez2024finite}, lattice Boltzmann~\cite{dunweg2007statistical}) 
%\IS{cite for each?}) \Garcia{Removed "spectral" since we couldn't find a good reference}
but in this introduction we focus on finite-volume schemes. 
After deriving the stochastic heat equation in Section~\ref{sec:SHE}, basic schemes are illustrated with results from a simple Python program (see Supplementary Materials) in Section~\ref{sec:numerics}.
The stochastic species diffusion equation plus a variety of other FHD equations are reviewed in Sections~\ref{sec:speciesFHD} and \ref{sec:menagerie}.
Finally, suggested exercises are listed in Section~\ref{sec:exercises}.

\section{Stochastic Heat Equation}\label{sec:SHE}

The methodology for defining stochastic fluxes that capture thermal fluctuations is based on ideas from statistical mechanics and nonequilibrium thermodynamics.  
From statistical mechanics~\cite{callen1985,Pathria_16}, we know that the probability density of fluctuations is of the form
$ %\[
%p \propto e^{-\Delta S / k_B}
p \propto \exp(-\Delta S / k_B)
$ %\]
where $\Delta S$ gives the fluctuation of entropy from the mean and $k_B$ is the Boltzmann constant. %\IS{define $k_B$}
In a given setting, an equation for the transport and production of entropy can be derived from the Gibbs relation using the equations describing the system (e.g., conservation of energy).
%\MarginPar{original was a bit terse.  not sure is this is the right way to say this but we need a transition here}
The resulting entropy production terms can be expressed in terms of the inner product of transport fluxes with gradients of the state variables.  
Specifically, the phenomenological laws of nonequilibrium thermodynamics characterize transport fluxes in terms of Onsager coefficients, $L$, multiplied by thermodynamic driving forces defined so that entropy production is given by the inner product of the thermodynamic forces with these fluxes~\cite{DeGroot_63,venerus2018modern}.
Finally, if the stochastic flux is taken to be a Gaussian random field with an amplitude given by
$ %\[
\sqrt{2 k_B L}
$, %\]
%then the resulting system predicts equilibrium fluctuations that match the predictions of statistical mechanics.
then the resulting system has equilibrium fluctuations that match the predictions of statistical mechanics.

As a first example of how this methodology is applied, we consider
the stochastic heat equation (SHE) for the transport of internal energy by thermal diffusion. 
This section presents a short derivation; for details, see~\cite{Zarate_07}.
We start with the transport equation for energy:
\begin{equation}
    \partial_ t (\rho e) = - \nabla \cdot \boldsymbol{Q},
    \label{eq:energy}
\end{equation}
where $\rho$ is the (constant) mass density and $e$ is the specific internal energy.
The heat flux, $\boldsymbol{Q}$, can be separated into deterministic and stochastic contributions (denoted by $\overline{\boldsymbol{Q}}$ and $\widetilde{\boldsymbol{Q}}$, respectively) and is written in Onsager form as
\begin{equation}
    \boldsymbol{Q} = \overline{\boldsymbol{Q}} + \widetilde{\boldsymbol{Q}} = L \boldsymbol{X} + \widetilde{\boldsymbol{Q}},
    \label{eq:ons}
\end{equation}
where $L$ is the Onsager coefficient and $\boldsymbol{X}$ is the thermodynamic force \cite{DeGroot_63,venerus2018modern}.
As noted above, the stochastic flux is given as a Gaussian random field with 
root-mean-square amplitude $\sqrt{2 k_B L}$.  Thus,
\begin{equation}
    \langle \widetilde{\boldsymbol{Q}}(\mathbf{r},t) \widetilde{\boldsymbol{Q}}(\mathbf{r}',t') \rangle 
    = 2 k_B L ~\delta(\mathbf{r}-\mathbf{r}') ~\delta(t - t') ~\mathcal{I},
\end{equation}
where $\mathcal{I}$ is the identity tensor and $\delta(\cdot)$ denotes the Dirac delta function.
Readers familiar with the Langevin equation for Brownian motion~\cite{Gardiner_85} will notice the similarities between this stochastic ordinary differential equation and the above stochastic PDE (SPDE).

From non-equilibrium thermodynamics~\cite{DeGroot_63, venerus2018modern}, 
the deterministic total rate of entropy change in a region $\Omega$ is
\begin{equation}
    \frac{d\overline{S}}{dt} = \int_\Omega \frac{\partial\overline{s}}{\partial t} \; d\mathbf{r} = \int_\Omega \boldsymbol{X} \cdot \boldsymbol{J} \; d\mathbf{r} - \left[\frac{\boldsymbol{J}}{T} \right]_{\partial\Omega},
    \label{eq:entrop}
\end{equation}
where $\boldsymbol{J}= L \boldsymbol{X}=\overline{\boldsymbol{Q}}$ is the thermodynamic flux and $T$ is the temperature.
The first term is the internal entropy production, and the second term is the entropy change due to heat flow at the boundary.
Using the Gibbs equation, $d\overline{u} = \rho d\overline{e} = T d\overline{s}$, and Eq.~\eqref{eq:energy} gives
\begin{equation}
    \int_\Omega \frac{\partial\overline{s}}{\partial t} \;d\mathbf{r} 
    = \int_\Omega \frac{\rho}{T} \frac{\partial\overline{e}}{\partial t}\; d\mathbf{r} 
    = -\int_\Omega \frac{1}{T} \nabla \cdot \overline{\boldsymbol{Q}} \; d\mathbf{r} 
    = \int_\Omega \frac{1}{T}  \nabla \cdot (\lambda\nabla T ) \; d\mathbf{r}.
    \label{eq:entrop2}
\end{equation}
The last equality uses the phenomenological Fourier law for heat flow,
$\overline{\boldsymbol{Q}} = - \lambda \nabla T$, where $\lambda$ is the (constant) thermal conductivity.
Applying Green's first identity gives
\begin{align}
    \int_\Omega \frac{1}{T} \nabla \cdot (\lambda\nabla T ) \; d\mathbf{r} 
    &= \left[ -\frac{\overline{\boldsymbol{Q}}}{T} \right]_{\partial\Omega}
     -\int_\Omega \nabla \left(\frac{1}{T}\right)  \cdot (\lambda\nabla T ) \; d\mathbf{r}.
     \label{eq:entrop3}
\end{align}
The entropy expressions \eqref{eq:entrop} and \eqref{eq:entrop2} can be equated, and so the chain rule gives
\begin{align}
    \int_\Omega \boldsymbol{X} \cdot \boldsymbol{J} \; d\mathbf{r}
    &=-\int_\Omega \nabla \left(\frac{1}{T}\right)  \cdot (\lambda\nabla T ) \; d\mathbf{r} 
    =\int_\Omega \nabla \left(\frac{1}{T}\right)  \cdot \left(\lambda T^2 \; \nabla \frac{1}{T} \right) \; d\mathbf{r}.
\end{align}
From this equality, 
$\boldsymbol{X}\cdot\boldsymbol{J} = \boldsymbol{X}\cdot(L\boldsymbol{X})$,
and Eq.~\eqref{eq:ons}, 
we can identify the thermodynamic force $\boldsymbol{X}$, the thermodynamic flux $\boldsymbol{J}$, and the Onsager coefficient $L$ to be~\cite{DeGroot_63, Garcia_22}
\begin{equation}
    \boldsymbol{X} = \nabla \left(\frac{1}{T}\right)\; \mathrm{,} \;\;\; 
    \boldsymbol{J} = \lambda T^2\; \nabla \left(\frac{1}{T}\right) = \overline{\boldsymbol{Q}}\;, \;\;\; \mathrm{and} \;\;\;
    L = \lambda T^2 ,
    \label{eq:ident}
\end{equation}
respectively.  
%(We note that from Eq. (\ref{eq:ident}) other ways of defining $L$ and $J$ are mathematically feasible; however, analysis of simple systems confirms that the decomposition given here is correct, see \cite{Garcia_22}).  
The noise correlation is then given by\footnote{Formally, the temperature in the noise intensity should be the deterministic value but in CFHD the instantaneous temperature is typically used.}
\begin{equation}
    \langle \widetilde{\boldsymbol{Q}}(\mathbf{r},t) \widetilde{\boldsymbol{Q}}(\mathbf{r}',t') \rangle
    = 2 k_B \lambda T^2 ~\delta( \mathbf{r} - \mathbf{r}') ~\delta (t - t') ~\mathcal{I}.
\end{equation}
Finally, by writing $e = c_V T$, where $c_V$ is the specific heat at constant volume, the SHE is given as
\begin{equation}
    \rho c_V \partial_t T = \nabla \cdot \left(\lambda \nabla T + \sqrt{2 k_B \lambda T^2}~\widetilde{\boldsymbol{Z}}\right),
\end{equation}
where $\widetilde{\boldsymbol{Z}}$ is a Gaussian white noise vector field uncorrelated in space and time, that is,
\begin{equation}
    \langle \widetilde{\boldsymbol{Z}}_s(\mathbf{r},t) \widetilde{\boldsymbol{Z}}_{s'}(\mathbf{r}',t')\rangle 
    = ~\delta( \mathbf{r} - \mathbf{r}') ~\delta (t - t') ~\delta^\mathrm{Kr}_{s,s'},\label{eq:GWN}
\end{equation}
where $s, s' \in \{ x, y, z \}$ denote Cartesian coordinate directions and $\delta^\mathrm{Kr}$ is Kronecker delta. %\MarginPar{Normalization of Z?} 

For simplicity we focus on the one-dimensional case (e.g., uniform rod) and write the SHE in compact form as
\begin{align}
    \partial_t T &= \partial_x \left(\kappa \partial_x T + \alpha T \tilde{Z}\right) 
                = \kappa \partial_x^2\, T + \alpha \partial_x\, (T \tilde{Z}),
    \label{eq:SHE}
\end{align}
where $\kappa = \lambda/\rho c_V$ and $\alpha = \sqrt{2 k_B \lambda}/\rho c_V$ are taken to be constants.
In the one-dimensional case, integrating over $y$ and $z$, the noise variance is
\begin{equation}
    \langle \tilde{Z}(x,t) \tilde{Z}(x',t')\rangle 
    = \frac{1}{\mathcal{A}}~\delta( x - x') ~\delta (t - t'),
\end{equation}
where $\mathcal{A}$ is the cross-sectional area of the system.

Assuming small fluctuations $\delta T = T - \bar{T}$ around the mean value $\bar{T}$, one can obtain from Eq.~\eqref{eq:SHE} the following linear SPDE with additive noise:
\begin{equation}
    \partial_t \delta T = \kappa \partial_x^2\, \delta T + \alpha \bar{T} \partial_x\, \tilde{Z}.
\end{equation}
This linearized SPDE has the form of an Ornstein--Uhlenbeck process that reaches an equilibrium distribution characterized by
$ %\begin{equation}
\lim_{t\rightarrow \infty} \langle \delta T(x,t) \rangle = 0
$ %\end{equation}
and
\begin{equation}
\lim_{t\rightarrow \infty}
\langle \delta T(x,t) \delta T (x',t) \rangle = \frac{\alpha^2 T_\mathrm{eq}^2}{2 \kappa \mathcal{A}} \delta(x-x') = \frac{k_B T_\mathrm{eq}^2}{\rho c_V \mathcal{A}} \delta(x-x'),
\label{eq:fdc}
\end{equation}
where $T_\mathrm{eq} = \bar{T}$ is the equilibrium thermodynamic temperature (see Supplementary Materials). %\MarginPar{John, Eqn (2.15) only for constant $\bar{T}$.}

While this linearized SHE can be mathematically rigorously defined with the use of distributions, 
the high irregularity of the stochastic flux makes interpreting the nonlinear SHE~\eqref{eq:SHE} mathematically ill-defined. 
In particular, the divergence of a random Gaussian field, which leads to distribution solutions in the linearized case, makes systems such as Eq.~\eqref{eq:SHE} too irregular to have a well-defined meaning \cite{gubinelli2015paracontrolled, hairer2014theory}; some type of regularization is required.
To obtain a mathematically tractable model, one needs to
introduce a high wavenumber cutoff that is of the order of the intermolecular distance \cite{nakano_2025_MD_FHD}.
In practice, we introduce this cutoff by discretizing the system using a finite-volume discretization with cells that are large enough to have at least $O(10)$ molecules per finite-volume cell, resulting in a finite-dimensional system of stochastic ordinary differential equations, as formulated in the next section.

\section{CFHD Schemes for SHE}\label{sec:numerics}

\begin{figure}
  \centering
    \includegraphics[width=0.90\textwidth]{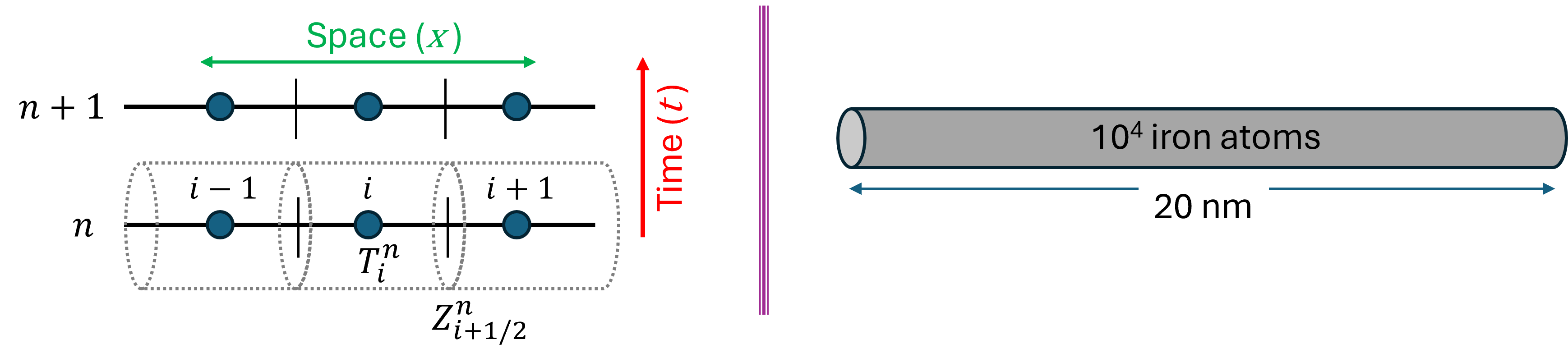}
    \caption{(left) Space and time discretization for SHE; (right) Physical system simulated by the \texttt{StochasticHeat} program ($\ell = 20~\mathrm{nm}$, $\mathcal{A} = 4~\mathrm{nm}^2$).}
    \label{fig:SHE_illustration}
\end{figure}

Equation~\eqref{eq:SHE} can be numerically integrated in time to produce sample trajectories for $T(x,t)$. 
As illustrated in Figure~\ref{fig:SHE_illustration}, we discretize space and time as $x_i = (i+\frac12) \Delta x$ and $t^n = n \Delta t$.
The temperature is given at cell centers, $T_i^n = T(x_i,t^n)$, for $i=0,\ldots,N-1$ where $N$ is the number of cells.   %}\MarginPar{Check}
We use a forward difference in time,
\begin{equation}
    \partial_t f  \Rightarrow \frac{f_i^{n+1} - f_i^n}{\Delta t},
\end{equation}
and centered differences in space,
\begin{equation}
    \partial_x f  \Rightarrow \frac{f_{i+\half}^n - f_{i-\half}^n}{\Delta x}\;, \qquad
    \partial_x^2 f  \Rightarrow \frac{f_{i+1}^n - 2 f_i^n + f_{i-1}^n}{\Delta x^2}\;.
\end{equation}
Note that for the spatial index, integer values are at cell centers, while half-integer values are at cell faces.

With this we have the forward Euler (FE) scheme:
\begin{equation}
    T_i^{n+1} = T_i^n 
    + \frac{\kappa \Delta t}{\Delta x^2} \left( T_{i+1}^n - 2 T_i^n  + T_{i-1}^n \right)
    + \frac{\alpha \Delta t}{\Delta x} 
    \left( T_{i+\half}^n \tilde{Z}_{i+\half}^n - T_{i-\half}^n \tilde{Z}_{i-\half}^n\right).
\end{equation}
For the temperature at a face we can use the arithmetic average, $T_{i+\half}^n = (T_{i+1}^n + T_i^n)/2$. 
Similarly, the predictor-corrector (PC) scheme is given as
\begin{align}
    T_i^* &= T_i^n 
    + \frac{\kappa \Delta t}{\Delta x^2} \left( T_{i+1}^n - 2 T_i^n + T_{i-1}^n \right)
    + \frac{\alpha \Delta t}{\Delta x} 
    \left( T_{i+\half}^n \tilde{Z}_{i+\half}^n - T_{i-\half}^n \tilde{Z}_{i-\half}^n\right), \\
    T_i^{n+1} &= \frac{1}{2} \Big[ T_i^n + T_i^* 
    + \frac{\kappa \Delta t}{\Delta x^2} \left( T_{i+1}^* - 2 T_i^* + T_{i-1}^* \right)
    + \frac{\alpha \Delta t}{\Delta x} 
    \left( T_{i+\half}^* \tilde{Z}_{i+\half}^n - T_{i-\half}^* \tilde{Z}_{i-\half}^n\right) \Big],
\end{align}
where $T_i^*$, computed in the predictor step, is used in the corrector step.\footnote{Note that the PC scheme is a trapezoidal scheme with constant noise over the time step. It is possible to develop improved discretizations of the stochastic flux using multiple random numbers; see \cite{Donev_CAMCOS_10}.}

The discretized noise has variance
\begin{equation}
    \left\langle \tilde{Z}_{i+\half}^n \tilde{Z}_{j+\half}^m \right\rangle 
    = \frac{1}{\mathcal{A}}~\frac{\delta^\mathrm{Kr}_{i,j}}{\Delta x} ~\frac{\delta^\mathrm{Kr}_{n,m}}{\Delta t} 
    = \frac{\delta^\mathrm{Kr}_{i,j}}{\Delta V}~\frac{\delta^\mathrm{Kr}_{n,m}}{\Delta t},
\end{equation}
where $\Delta V$ is the volume of a grid cell.
Numerically this noise is generated as
\begin{equation}
    \tilde{Z}_{i+\half}^n = \frac{1}{\sqrt{\Delta V \Delta t}}~\mathfrak{N}_{i+\half}^n,
\end{equation}
where $\mathfrak{N}$ are independent, normally distributed (Gaussian) psuedo-random numbers {with zero mean and unit variance}.

The discretization described above requires some minor modifications for Dirichlet boundary conditions that reflect the boundary conditions being prescribed at cell faces.
For the left and right boundary conditions, $T_\mathrm{L} = T(x=0)$ and $T_\mathrm{R} = T(x=L)$, the second derivative $\partial_x^2 T$ is approximated as
\begin{equation}
\frac{1}{\Delta x^2} \left( T_1 - 3 T_0  + 2T_\mathrm{L} \right) 
\;\; \mathrm{and} \;\;
\frac{1}{\Delta x^2} \left( T_{N-2} - 3 T_{N-1}  +2 T_\mathrm{R} \right) \; ,
\end{equation}
% From StochasticHeat program:
%   Determ[0] = coeffDetFE * ( T[ 1 ] + 2 * T_Left - 3*T[ 0 ] )  # Left Dirichlet BC
%   Determ[Ncells-1] = coeffDetFE * ( 2*T_Right + T[Ncells-2] - 3*T[Ncells-1] )  # Right Dirichlet BC
% \begin{equation}
% \frac{1}{\Delta x^2} \left( T_{i+1} - 3 T_i  + 2T_B^L \right) 
% \;\; \mathrm{and} \;\;
% \frac{1}{\Delta x^2} \left( T_{i-1} - 3 T_i  +2 T_B^R \right) \; ,
% \end{equation}
in cells $0$ and $N-1$, respectively.  
To ensure that the discrete system captures the correct variance at equilibrium, the noise term at the leftmost face $(x=0)$ is modified to $\sqrt{2}\; T_\mathrm{L}\; \tilde{Z}_{-\frac12}$ and similarly to $\sqrt{2}\; T_\mathrm{R}\; \tilde{Z}_{N-\frac12}$ for the rightmost face $(x=L)$.  %}\MarginPar{Check}

The Python program, \texttt{StochasticHeat}, (see Supplementary Materials) illustrates the numerical methods described in this section for the SHE.
The simulation computes trajectories, $T_i^n$, discretized in space and time for the one-dimensional SHE~\eqref{eq:SHE} and analyzes their statistical properties (e.g., variance of temperature). 
The reader is encouraged to download and run the program, as results are obtained in about 5 minutes on a laptop.
An alternative is to add the stochastic heat flux, as described above, to an existing program that solves the heat equation by a finite-volume method.

The \texttt{StochasticHeat} program has various global options and parameters:
\begin{itemize}%[noitemsep,topsep=0pt] 
    \item Periodic or Dirichlet boundary conditions,
    \item Thermodynamic equilibrium or constant temperature gradient,
    \item Initialize with or without temperature perturbations,
    \item Run simulation with or without thermal fluctuations,
    \item Use the forward Euler (FE) or predictor-corrector (PC) scheme,
    \item Number of grid cells, $N$,
    \item Number of time steps, $M$.
\end{itemize}
The physical system is a small iron rod ($\approx 10^4$ atoms) of length $\ell = 20~\mathrm{nm}$ and cross-sectional area $\mathcal{A} = 4~\mathrm{nm}^2$  (see Fig.~\ref{fig:SHE_illustration}). 
For iron $\rho = 7870~\mathrm{kg/m}^3$, $c_V = 450~\mathrm{J/kg\cdot K}$, and $\lambda = 70~\mathrm{W/m\cdot K}$. 
For numerical stability $\Delta t = a \Delta x^2 / 2 \kappa$ with $a \leq 1$; the \texttt{StochasticHeat} program sets $a = 1/10$.
In the equilibrium cases $T_\mathrm{eq} = 300~\mathrm{K}$ while in the temperature gradient case the left and right boundaries are fixed at $100~\mathrm{K}$ and $500~\mathrm{K}$, respectively.
The simulation results in this section are from runs using $N=32$~grid cells, running for $M = 2\times 10^6$ (equilibrium cases) or $M = 8\times 10^6$ (temperature gradient case)~time steps.

The program computes time-averaged quantities; for example, the average temperature in a cell is
\begin{equation}
    \langle T_i \rangle = \frac{1}{M_\mathrm{s}}~ \sum_{n = n_\mathrm{r}+1}^{n_\mathrm{r} + M_\mathrm{s}} T_i^n,
\end{equation}
where $M_\mathrm{s}$ is the number of statistical samples; the system ``relaxes'' for the first $n_\mathrm{r}$ time steps before sampling begins. 

\begin{figure}
  \centering
    \includegraphics[width=0.45\textwidth]{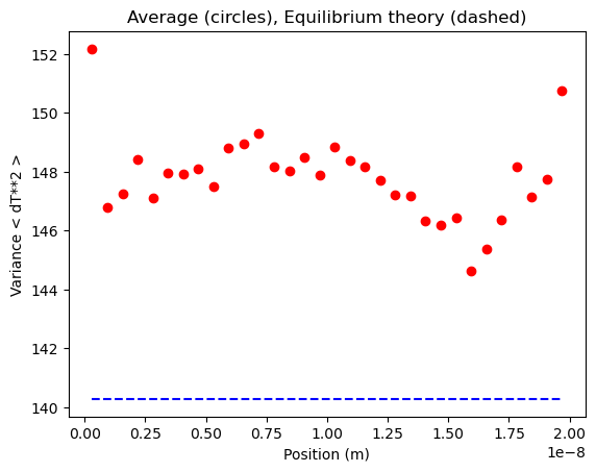}
    \includegraphics[width=0.45\textwidth]{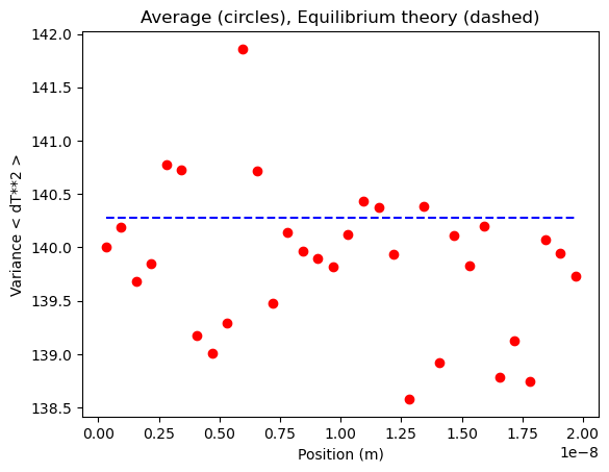}
    \caption{Temperature variance $\langle \delta T_i^2 \rangle$ versus $x_i$ at equilibrium with Dirichlet boundary conditions for (left) FE scheme; (right) PC scheme.
    Theory (dashed line) is given by Eq.~\eqref{eq:varcorrT_EQ}.}
    \label{fig:VarEquilib}
\end{figure}

\begin{figure}
  \centering
    \includegraphics[width=0.45\textwidth]{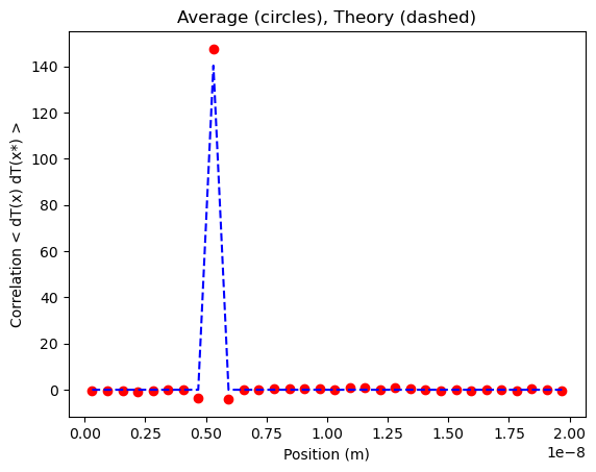}
    \includegraphics[width=0.45\textwidth]{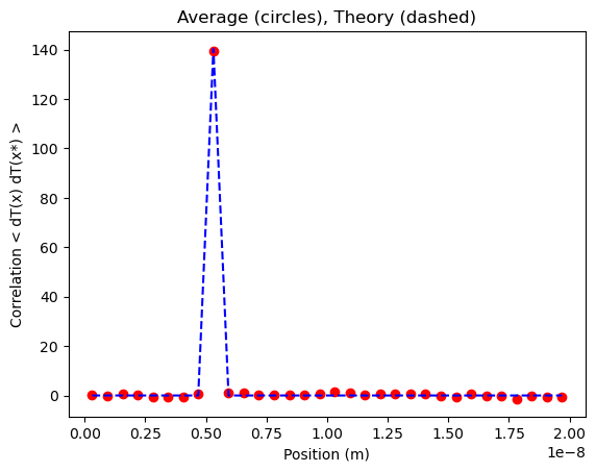}
    \caption{Temperature correlation $\langle \delta T_i \delta T_j \rangle$ versus $x_i$ at equilibrium with Dirichlet boundary conditions with $x_j = \ell/4$ for (left) FE scheme; (right) PC scheme.
    Theory line given by Eq.~\eqref{eq:varcorrT_EQ}.}
    \label{fig:CorrEquilib}
\end{figure}

\begin{figure}
  \centering
    \includegraphics[width=0.45\textwidth]{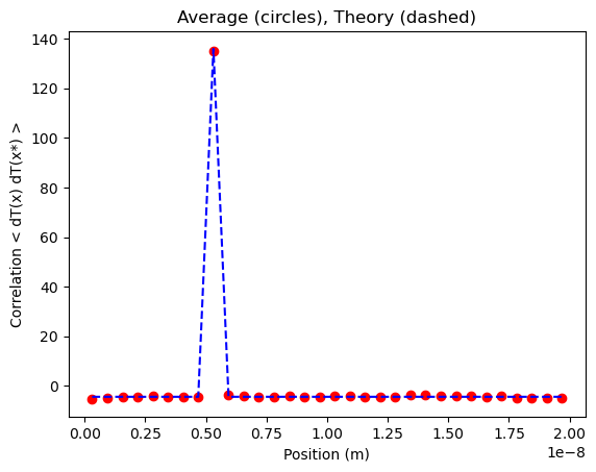}
    \includegraphics[width=0.45\textwidth]{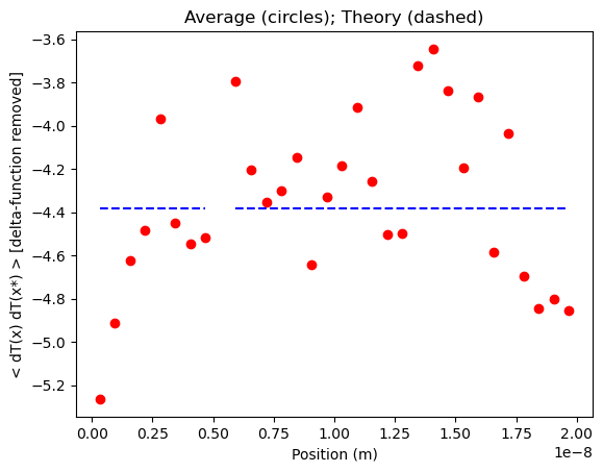}
    \caption{Temperature correlation $\langle \delta T_i \delta T_j \rangle$ versus $x_i$ at equilibrium with periodic boundary conditions using the PC scheme.
    Results are plotted both with (left) and without (right) the point at $x_j = \ell/4$.}
    \label{fig:VarPeriodic}
\end{figure}

At thermodynamic equilibrium the variance and static correlation of temperature fluctuations ($\delta T_i^n = T_i^n - \langle T_i\rangle$) are predicted by statistical mechanics to be~\cite{Pathria_16}
\begin{equation}
    \langle \delta T_i^2 \rangle = \frac{k_B T_\mathrm{eq}^2}{\rho c_V \Delta V}
    \qquad\mathrm{and}\qquad
    \langle \delta T_i \delta T_j \rangle = \langle \delta T_i^2 \rangle ~ \delta^\mathrm{Kr}_{i,j}
    \label{eq:varcorrT_EQ}
\end{equation}
with the average temperature, $\langle T_i \rangle$, equal to the equilibrium temperature, $T_\mathrm{eq}$.
Equation~\eqref{eq:varcorrT_EQ} is the discrete analog of Eq.~\eqref{eq:fdc}.
Figures~\ref{fig:VarEquilib} and \ref{fig:CorrEquilib} show that the equilibrium simulation results for Dirichlet boundaries (i.e., fixed temperature at the boundaries) are in good agreement with theory, especially for the PC scheme.
Figure~\ref{fig:VarPeriodic} shows that temperature correlations are slightly different for periodic boundary conditions since, due to the conservation of energy, $\sum_i \rho c_V \langle \delta T_i \rangle = 0$, the correlation $\langle \delta T_i \delta T_j \rangle$ for $i\neq j$ is shifted by an additive constant such that $\sum_i \langle \delta T_i \delta T_j \rangle =\langle \delta T_j^2 \rangle + \sum_{i\neq j} \langle \delta T_i \delta T_j \rangle = 0$. This is a common feature in systems with conserved variables, such as density fluctuations in closed systems.

\begin{figure}
  \centering
    \includegraphics[width=0.45\textwidth]{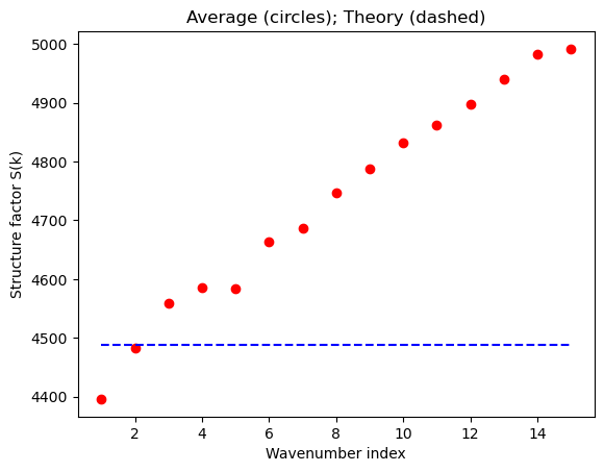}
    \includegraphics[width=0.45\textwidth]{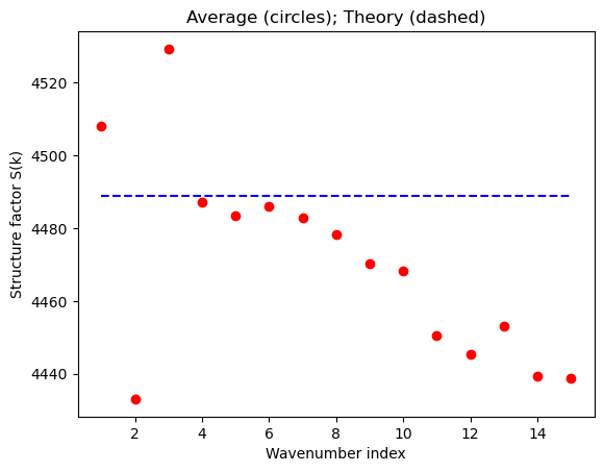}
    \caption{Static structure factor, $S_k$, versus $k$ at equilibrium for (left) FE scheme; (right) PC scheme.
    Theory line given by Eq.~\eqref{eq:SkTheory}.}
    \label{fig:StructFactor}
\end{figure}

The assessment of numerical schemes in CFHD is best done by measuring the static structure factor for fluctuations. 
In simple terms, the structure factor is the time-averaged correlations of the state variables, in this case temperature, in Fourier space, and is expressed as a function of wavenumber.
For the SHE, we define the unnormalized structure factor, $S_k = \langle \hat{T}_k \hat{T}^*_k \rangle$, where
\begin{equation}
    \hat{T}_k = \sum_{j=0}^{N-1} T_j \exp (- 2\pi \mathrm{i}\, j k / N)
\end{equation}
is the discrete Fourier transform with wavenumber index $k = 0, 1, \ldots, N-1$, and $\hat{T}^*_k$ is its complex conjugate.\footnote{In CFHD the structure factor is often normalized to unity using the equilibrium variance.}
From Eq.~\eqref{eq:varcorrT_EQ}, at thermodynamic equilibrium, the discretized structure factor is
\begin{equation}
    S_k  = \frac{k_B T_\mathrm{eq}^2}{\rho c_V}~N.
    \label{eq:SkTheory}
\end{equation}
Figure~\ref{fig:StructFactor} shows that the FE scheme has significant errors for large $k$ while the PC scheme is more accurate for all $k$. 
An analysis of discretization errors for various numerical schemes is presented in \cite{Donev_CAMCOS_10}.

\begin{figure}
  \centering
    \includegraphics[width=0.45\textwidth]{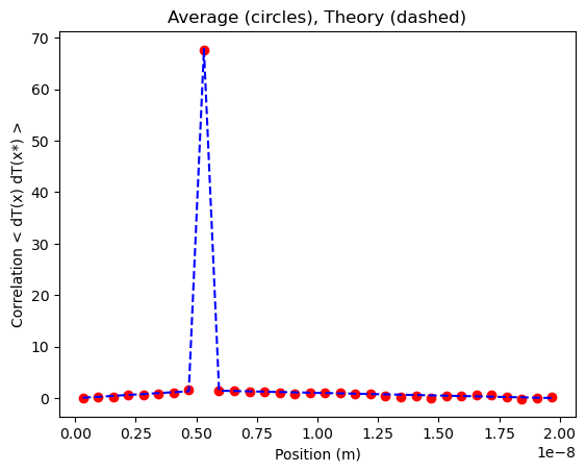}
    \includegraphics[width=0.45\textwidth]{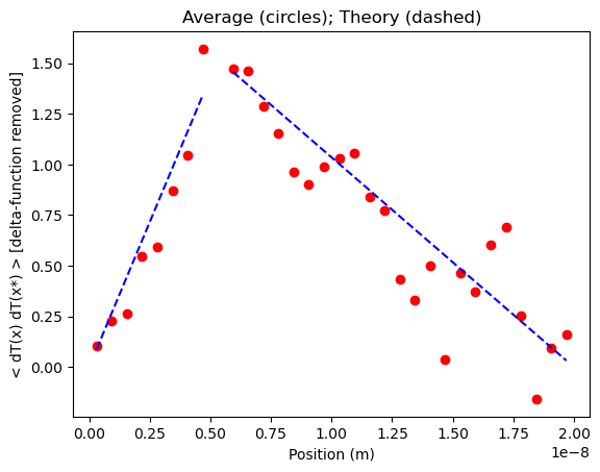}
    \caption{Temperature correlation $\langle \delta T_i \delta T_j \rangle$ versus $x_i$ 
    for linear temperature gradient imposed by Dirichlet boundary conditions using the PC scheme. Results are plotted with (left) and without (right) the point at $x_j = \ell/4$; theory line given by Eq.~\eqref{eq:TcorrNEQ}.}
    \label{fig:TcorrNEQ}
\end{figure}

Finally, we consider a non-equilibrium scenario, specifically a linear temperature gradient imposed by Dirichlet boundary conditions.
In this case there is a weak, long-range correlation of temperature fluctuations~\cite{Garcia_JSP_87}:
\begin{equation}
    \langle \delta T_i \delta T_j \rangle = 
    \frac{k_B \langle T_i \rangle^2}{\rho c_V \Delta V} \delta^\mathrm{Kr}_{i,j}
    + \frac{k_B (\nabla T)^2}{\rho c_V \mathcal{A} \ell} %~ x_< (L - x_>)
    \times \left\{ \begin{array}{cc}
   x_i (\ell - x_j) &  \mathrm{if}~ x_i < x_j,\\
   x_j (\ell - x_i) & \mathrm{otherwise}.
\end{array}
\right.
\label{eq:TcorrNEQ}
\end{equation}
Figure \ref{fig:TcorrNEQ} shows that the simulation results are in good agreement with this theoretical result.

\section{Stochastic Species Diffusion Equation}\label{sec:speciesFHD}

In FHD there are several stochastic diffusion equations closely related to SHE. 
For example, the diffusion of species mass is described by a similar SPDE with the main difference being the functional form of the stochastic flux. 
As before we start with
\begin{equation}
    \partial_t (\rho c) = - \nabla \cdot \boldsymbol{F},
\end{equation}
where $c$ is the species mass fraction and the species flux is
\begin{equation}
    \boldsymbol{F} = \overline{\boldsymbol{F}} + \widetilde{\boldsymbol{F}} = \mathcal{L} \boldsymbol{\mathcal{X}} + \widetilde{\boldsymbol{F}}.
\end{equation}
Assuming that the system is isothermal (i.e., no Soret effect), irreversible thermodynamics tells us that the thermodynamic force is~\cite{DeGroot_63, Garcia_22}
\begin{equation}
    \boldsymbol{\mathcal{X}} =  -\nabla \left(\frac{\mu}{T}\right)  = -\frac{\nabla \mu}{T},
\end{equation}
where $\mu$ is the chemical potential.
For ideal solutions,
\begin{equation}
\mu = \mu_0(T) + \frac{k_B T}{m} \ln c 
\qquad\mathrm{so}\qquad
\nabla \mu = \frac{k_B T}{mc}~ \nabla c ,
\end{equation}
where $\mu_0(T)$ is chemical potential of pure substance (i.e., $c=1$) at temperature $T$ and $m$ is the particle mass.
Since the phenomenological law for species flow is Fick's law,
$\overline{\boldsymbol{F}} = - \rho D \nabla c$,
the Onsager coefficient is
\begin{equation}
    %\mathcal{L} = {m \rho D\, c}/{k_B}\; .
    \mathcal{L} = \frac{m \rho D\, c}{k_B}.
\end{equation}
The noise correlation is
\begin{equation}
    \langle \widetilde{\boldsymbol{F}}(\mathbf{r},t) \widetilde{\boldsymbol{F}}(\mathbf{r}',t') \rangle
    = 2 m \rho D \, c ~\delta( \mathbf{r} - \mathbf{r}') ~\delta (t - t') ~\mathcal{I},
\end{equation}
so the stochastic species diffusion equation is
\begin{equation}
    \partial_t\, c = \nabla \cdot \left(D \nabla c 
    + \sqrt{\frac{2 D \, c}{n_0}}~\widetilde{\boldsymbol{Z}} \right),
    \label{eq:SSDE}
\end{equation}
where $n_0 = \rho/m$ is the number density of the pure substance state.

This SPDE can also be written in terms of species number density, $n = c\, n_0$, as
\begin{equation}
    \partial_t\, n = \nabla \cdot \left(D \nabla n 
    + \sqrt{2 D \, n}~\widetilde{\boldsymbol{Z}}\right).
    \label{eq:DK}
\end{equation}
In this form it is known as the Dean--Kawasaki equation \cite{dean1996langevin,kawasaki1994stochastic}, which can also be derived by coarse-graining the discrete random walk model for diffusion. 
%\CK{References?}\MarginPar{I went back to the original dean and kawasaki papers.  that seem ok? -- JBB}\Garcia{John, what's a good reference to add here?}
Finally, recall that the SHE is
\begin{equation}
     \partial_t\, T = \nabla \cdot \left(\kappa \nabla T 
    + \sqrt{\frac{2 \kappa k_B \,T^2}{\rho c_V}}~\widetilde{\boldsymbol{Z}} \right).
    \label{eq:SHE2}
\end{equation}
It is important to note that the deterministic parts of the two equations above are functionally similar, while the stochastic parts are distinctly different.
Specifically, the stochastic term goes as $\sqrt{n}$ in Eq.~\eqref{eq:DK} and as $T$ in Eq.~\eqref{eq:SHE2}.

\section{A Menagerie of FHD Equations}\label{sec:menagerie}

We have so far focused on the SHE and stochastic species diffusion equation due to the pedagogical benefit of their simplicity.
However, various FHD models have been formulated for a wide range of physical systems.  
In this section, we briefly discuss a number of different models.
The general form of the FHD equations for a set of state variables $\boldsymbol{U}$ is
\begin{equation}
    \partial_t \boldsymbol{U} = - \nabla \cdot \left[ \boldsymbol{\mathcal{F}}_H(\boldsymbol{U}) + \boldsymbol{\mathcal{F}}_D(\boldsymbol{U}) - \boldsymbol{\mathcal{N}}(\boldsymbol{U})\widetilde{\boldsymbol{Z}}\right],
    \label{eq:genFHD}
\end{equation}
where $\boldsymbol{\mathcal{F}}_H$ and $\boldsymbol{\mathcal{F}}_D$ are the hyperbolic and diffusive fluxes, respectively, and $\boldsymbol{\mathcal{N}}$ denotes the noise intensity.\footnote{Since $\widetilde{\boldsymbol{Z}}$ and $-\widetilde{\boldsymbol{Z}}$ have the identical statistical properties, it does not in fact matter whether a positive or negative sign is used for the term $\boldsymbol{\mathcal{N}}(\boldsymbol{U})\widetilde{\boldsymbol{Z}}$ in Eq.~\eqref{eq:genFHD}.}
For example, for Eq.~\eqref{eq:DK}, $\boldsymbol{U} = n$, $\boldsymbol{\mathcal{F}}_H = 0$, $\boldsymbol{\mathcal{F}}_D = -D\nabla n$, and $\boldsymbol{\mathcal{N}} = \sqrt{2 D n}$.
In general, the fluctuation-dissipation theorem relates $\boldsymbol{\mathcal{F}}_D$ to $\boldsymbol{\mathcal{N}}$ (see Supplementary Materials) and tells us that the stochastic term is independent of the hyperbolic term.

Below is an outline of other FHD equations including references describing finite-volume schemes for solving them.
\begin{itemize}
    \item Stochastic Burger's equation~\cite{Bell_JCP_07}:
    \begin{equation}
        \partial_t\, c = -\partial_x\, \left( a c (1-c) - D\, \partial_x c + \sqrt{2 D c (1-c)}\, \tilde{Z} \right).
    \end{equation}
    Note that for $c \ll 1$ this resembles the stochastic species diffusion equation with the addition of a hyperbolic flux, $\boldsymbol{F}_H = a c (1-c)$.

    \item Stochastic ``train model'' equations~\cite{Alexander_JCP_05,Garcia_AJP_96}:
    \begin{align}
        \partial_t\, \rho &= - \partial_x \left( -D \partial_x \rho + \sqrt{2 m D \rho }\, \tilde{Z}\right), \\
        \partial_t\, (\rho u) &= - \partial_x \left( -D \partial_x (\rho u) + \sqrt{2 m D \rho u^2 }\, \tilde{Z}\right).
    \end{align}
    These equations describe a simple model for the transport of momentum density, $\rho u$, in a fluid.

    \item FHD for reaction-diffusion systems~\cite{Kim_JCP_17}:\\
    The reaction-diffusion system has $N_s$ species diffusing and undergoing $N_r$ reactions.  By denoting the number density of species $s$ by $n_s$, the equations of fluctuating reaction-diffusion can be written formally as the SPDEs:
    \begin{equation}
        \partial_t\, n_s = \nabla\cdot\left(D_s\nabla n_s + \sqrt{2D_s n_s}\boldsymbol{\mathcal{Z}}_s^{(D)}\right) + \sum_{r=1}^{N_r}\nu_{sr}\left(a_r(\boldsymbol{n}) + \sqrt{a_r(\boldsymbol{n})}\mathcal{Z}_r^{(R)}\right),
    \end{equation}
   where $D_s$ is the diffusion coefficient of species $s$, $\boldsymbol{n} = 
   (n_1, \ldots, n_{N_s})$, $a_r(\boldsymbol{n})$ 
   is the propensity function indicating the rate of reaction $r$, and $\nu_{sr}$ is the stoichiometric coefficient of species $s$ in reaction $r$.

    \item Stochastic compressible Navier--Stokes equations (single-species fluid)~\cite{Bell_PRE_07,Srivastava_PRE_23}:
    \begin{align}
    \partial_t\, \rho &= - \nabla \cdot \left( \rho \boldsymbol{u} \right), \label{eq:Continuity}\\
    \partial_t\, ( \rho \boldsymbol{u} ) &=  - \nabla \cdot \left[ ( \rho \boldsymbol{u} \otimes \boldsymbol{u} + p\mathbb{I} )  + \overline{\Pi} + \widetilde{\Pi} \right],\\
    \partial_t\, ( \rho E ) &= - \nabla \cdot \left[ \left(\rho E + p\right) \boldsymbol{u} + \overline{\boldsymbol{Q}} + \widetilde{\boldsymbol{Q}}  + \left( \overline{\Pi} + \widetilde{\Pi} \right) \cdot  \boldsymbol{u} \right],
    \label{eq:CompressibleE}
    \end{align}
    where $\boldsymbol{u}$, $p$, and $\Pi = \overline{\Pi} + \widetilde{\Pi}$ are fluid velocity, pressure, and stress tensor; total specific energy is $E = e + \frac12 u^2$. Note that Eq.~\eqref{eq:CompressibleE} reduces to the stochastic heat equation when $\boldsymbol{u}=0$. See references for explicit expression of the deterministic and stochastic fluxes.

    \item Stochastic compressible Navier--Stokes equations (multi-species fluid)~\cite{Balakrishnan_PRE_14, Srivastava_PRE_23}:\\
    Same as above but replace Eq.~\eqref{eq:Continuity} with
    \begin{equation}
        \partial_t\, \rho_s = - \nabla \cdot ( \rho_s \boldsymbol{u} 
        + \overline{\boldsymbol{F}}_s + \widetilde{\boldsymbol{F}}_s ),
        \label{eq:SpeciesFHD}
    \end{equation}
    where $\rho_s$ is the mass density for species $s$ and $\rho = \sum_s \rho_s$. For ideal mixtures the diffusive flux, $\boldsymbol{F}_s = \overline{\boldsymbol{F}}_s + \widetilde{\boldsymbol{F}}_s$, is similar to that of the species diffusion equation described in the previous section, where the diffusion coefficient is replaced by a matrix of diffusion coefficients and where species diffusion and the heat flux are modified to include Soret and Dufour effects, respectively.

    \item Stochastic incompressible / low-Mach FHD~\cite{balboa2012staggered, Donev_PoF_15, Donev_CAMCOS_14, nonaka2015low, ATZBERGER20071255}:\\
    The multi-species methodology can be extended to model isothermal mixtures of miscible incompressible liquids.
    Incompressibility of the fluids leads to a constrained evolution given by
    \begin{align}
    \partial_t\, ( \rho \boldsymbol{u} ) &=  - \nabla \cdot \left[  \rho \boldsymbol{u} \otimes \boldsymbol{u} +  \overline{\Pi} + \widetilde{\Pi} \right] - \nabla \pi,\\
    \nabla \cdot \boldsymbol{u} &= - \nabla \cdot \left (\sum_s \frac{ \overline{\boldsymbol{F}}_s + \widetilde{\boldsymbol{F}}_s}{\bar{\rho}_s} \right ).
    \label{eq:lowMach}
    \end{align}
    and Eq.~\eqref{eq:SpeciesFHD}.
    Here, $\pi$ is a perturbational pressure and $\bar{\rho}_s$ is the pure component density for species $s$.
    The resulting system retains the influence of fluctuations and is considered to be more computationally efficient, particularly for liquids with a high sound speed.

    \item FHD with chemical reactions~\cite{Bhattacharjee_JCP_15, Kim_JCP_18}:\\
    Reactive source terms containing the deterministic and stochastic chemical production rates, $\overline{\Omega}_s$ and $\widetilde{\Omega}_s$, respectively, are added to the right-hand side of Eq.~\eqref{eq:SpeciesFHD}.
    These can be formulated from the chemical Langevin equation~\cite{Gillespie_JCP_00} and included in either the compressible or incompressible hydrodynamic equations.
       
    \item Stochastic Poisson--Nernst--Planck equations~\cite{Donev_PRF_19, Peraud_PRF_16}:\\
    The incompressible version of CFHD for multi-species fluids can model electrolyte solutions by including charged species (ions).
    The chemical potential becomes the electrochemical potential with the electrical mobility given by the Nernst--Einstein relation and the diffusive flux by the Nernst--Planck equation.
    For scales comparable to the Debye length, the electric field is obtained by solving the Poisson equation.
    At scales larger than the Debye length, the fluid is electro-neutral.
    Electro-neutrality can be imposed as a constraint by solving a modified elliptic equation to compute the electric potential.

    \item FHD for multi-phase fluids~\cite{barker2023fluctuating, Chaudhri_PRE_14, klymko2020low, LAZARIDIS20171431, shang2011fluctuating}:\\
    By adding non-local gradient terms to the free energy, one can incorporate interfacial tension in CFHD model to treat multi-phase systems.  
    This can be done for both single component systems (e.g., liquid/vapor) and for multicomponent systems (e.g., oil/water). 
    For multicomponent systems, the low Mach number version of the methodology has also been derived.  
    These types of systems introduce additional higher order operators into the momentum equations and, for multicomponent systems, the species transport equations.  See the references for specific forms of these equations.

\end{itemize}
\noindent A repository of CFHD codes written using the \texttt{AMReX} framework \cite{AMReX:IJHPCA} is available at \url{https://github.com/AMReX-FHD}.

\section{Suggested Exercises}\label{sec:exercises}

Listed below are exercises suitable for homework or projects in a numerical PDE course. For each exercise you can either start with the \texttt{StochasticHeat} program and modify it or you can start from scratch and write your own program using the concepts presented above.

\begin{enumerate}

    \item  For the stochastic heat equation, measure the time-correlation $\langle \delta T_i^{n+m} \delta T_i^n \rangle$ and plot it as a function of $m$. How does this time-correlation depend on the thermal conductivity, $\lambda$?

    \item Verify numerically that the error in $S_k$ is $O(\kappa \Delta t\, k^2)$ for the forward Euler scheme and is $O(\kappa^2 \Delta t^2 k^4)$ for the predictor-corrector scheme.

    \item Write a program to solve the stochastic heat equation using an implicit scheme, such as backward Euler or Crank--Nicolson. Report on stability and accuracy (based on structure factor) as a function of timestep size.

    \item Consider a system, similar to the rod simulated by the \texttt{StochasticHeat} program, which is iron from $x=0$ to $\ell/2$ and aluminum $x=\ell/2$ to $\ell$. Do the variance and correlation of temperature fluctuations change when the system is at equilibrium? What about when it has a temperature gradient?

    \item Write a program to compute the stochastic species diffusion equation for the number density, see Eq.~\eqref{eq:DK}. Compare the simulation results with the theoretical equilibrium prediction, $\langle \delta n_i \delta n_j \rangle = \delta_{i,j}^\mathrm{Kr} \langle n_i \rangle / \Delta V$.
    Are the correlations different when the system has a number density gradient?

    \item Write a program to compute the stochastic Burgers' equation (see~\cite{Bell_JCP_07} for guidance). 

    \item Write a program to compute the stochastic train model (see \cite{Alexander_JCP_05,Garcia_AJP_96} for guidance).

\end{enumerate}

\section{Conclusions}\label{sec:conclusions}

At macroscopic scales, fluid dynamics is governed by partial differential equations
that characterize the behavior of a fluid in terms of smoothly evolving fields
that represent  density, momentum, and other characteristics of the fluid.  
At atomic scales, fluids are discrete systems composed of individual molecules whose
dynamics are governed by complex interaction potentials.
There has been extensive development of numerical methodology for simulating fluids in both the macroscopic and the molecular scales.
However, at intermediate mesoscopic scales, neither macroscopic nor atomistic approaches are appropriate.
At the mesoscale, a macroscopic formulation is not applicable because it fails to include the effects of fluctuations, whereas although atomistic approaches capture fluctuations, they are prohibitively expensive.
Fluctuating hydrodynamics provides an efficient, cost-effective approach to modeling the effect of fluctuations in mesoscale systems. 
This capability to capture the impact of fluctuations on mesoscale behavior and the implications of that behavior on macroscale dynamics is critical to quantifying the emergent behavior of systems designed at the nanoscale.

\section*{Acknowledgments}
The authors acknowledge support from the US Department of Energy, Office of Science, Office of Advanced Scientific Computing Research, Applied Mathematics Program under contract no.\ DE-AC02-05CH11231.

\bibliographystyle{siamplain}
\bibliography{MAIN_Bib}  % Produces the bibliography via BibTeX.

\end{document}

\maketitle

\section{Fluctuation-dissipation balance in Ornstein--Uhlenbeck processes}

Linearized versions of fluctuating hydrodynamics models are stochastic processes of the form,
\begin{equation}
    %du = -\LL u\; dt + \BB \; dW,
    du = -\LL u\; dt + \BB \; dW,\quad u(0) = u_0,
    \label{eq:OUP}
\end{equation}
%where $\LL$ and $\BB$ are linear operators and $dW$ is a cylindrical Wiener process (independent vector random processes in finite dimensions).  
where $\LL$ and $\BB$ are linear operators and $W$ is a cylindrical Wiener process (consisting of independent vector-valued random processes in finite dimensions).  
%\MarginPar{Using $\mathcal{L}$ to avoid confusion with Onsager coefficient}
Stochastic processes of this form are referred to as Ornstein--Uhlenbeck processes (OUP).
For an OUP process, $\BB \; dW$ adds noise to the solution that is damped by the deterministic term.
%\footnote{Here we assume that the real part of the spectrum of $L$ is positive.}
%\CK{The assumption in the footnote could be in the main text because it is used later in the derivation.}
For simplicity of exposition, we here assume that the real part of the spectrum of $\LL$ is strictly positive.
%A well-known result for these systems is that their long-time behavior converges to an equilibrium distribution, see~\cite{Gardiner_85}.
A well-known result for these systems is that their long-time behavior converges to an equilibrium distribution~\cite{Gardiner_85}.

% Here we present a simple derivation of the covariance of the equilibrium distribution.
Here we present a simple derivation for the covariance of the equilibrium distribution.
%We want to characterize the \Add{equilibrium} covariance, which is the expectation of $S \equiv \langle u(t),u^*(t) \rangle$ as $t \rightarrow \infty$. 
In other words, we want to characterize the equilibrium covariance, which is given as
\begin{equation}
    S = \lim_{t \rightarrow \infty} \langle u(t) u^*(t) \rangle,
\end{equation}
where * denotes adjoint (or conjugate transpose).

From Duhamel's principle, we can write the solution of Eq.~\eqref{eq:OUP} as
\begin{equation}
    u(t) = e^{-\LL t} u_0 + \int_0^t e^{\LL(s-t)} \BB \; dW(s).
\end{equation}
%From the assumptions on $\LL$, the first term in the solution vanishes as $t \rightarrow \infty$ and the mean of the solution is zero.
From the assumption on $\LL$, the first term in the solution vanishes as $t \rightarrow \infty$ and the mean of the solution also becomes zero.
%Thus\Add{, we have}
We thus have
%\CK{I replaced large $(\cdot,\cdot)$ by large $\langle\cdot,\cdot\rangle$ in the following equation because it seems that $\langle\cdot,\cdot\rangle$ are used as (the expectation of) inner product in this SM. Note, however, that in the main manuscript we use $\langle\cdot\rangle$ as the expectation of a quantity (or a tensor). So I deleted commas in the main manuscript, e.g., $\langle \delta T(x,t), \delta T (x',t) \rangle \rightarrow \langle \delta T(x,t) \delta T (x',t) \rangle$.
%One more thing: if $\langle\cdot,\cdot\rangle$ here means inner product, * is in fact not needed (i.e., $\langle a,b\rangle = \mathbb{E}[ab^*]$). Suggestion: We could simply omit commas also here in this SM and use $\langle\cdot\rangle$ as expectation.}
\begin{equation}
% \lim_{t\rightarrow\infty}  \langle u(t),u^*(t) \rangle = \lim_{t\rightarrow\infty} 
% \left\langle  \int_0^t e^{\LL(s-t)} \BB \; dW(s) \;,\;
% \left ( \int_0^t e^{\LL(s'-t)} \BB \; dW(s')\right )^* \right\rangle.
    S = \lim_{t\rightarrow\infty} 
    \left\langle  \int_0^t e^{\LL(s-t)} \BB \; dW(s) \:
    \left ( \int_0^t e^{\LL(s'-t)} \BB \; dW(s')\right )^* \right\rangle.
\end{equation}
%Since the noise is uncorrelated in time \CK{Please check the updated equation of (SM1.4)}
%\CK{One more thing: Since $L$ and $B$ are not in boldface, $I$ could also be not in boldface.}
Since the noise is uncorrelated in time,
\begin{equation}
%\langle dW(s), dW(s') \rangle = \mathbf{I}\; ds
    \langle dW(s)\; dW^*(s') \rangle = \mathcal{I}\: \delta(s-s')\: ds\: ds',
\end{equation}
%\Add{where $\mathbf{I}$ is the identity tensor.}
where $\mathcal{I}$ is the identity tensor.
%\NewOld{And so we have}{so the} \CK{Please double check whether the minus sign is expected in $\exp(-\LL^*(s-t))$.}
We thus have
\begin{equation}
% \lim_{t\rightarrow\infty}  \langle u(t),u^*(t) \rangle = \lim_{t\rightarrow\infty} 
%  \int_0^t e^{\LL(s-t)} \BB \BB^*
% e^{-\LL^* (s-t)} ds.
    S = \lim_{t\rightarrow\infty} \int_0^t e^{\LL(s-t)} \BB \BB^* e^{\LL^* (s-t)} ds.
\end{equation}
% We then observe that
Since $S^* = S$, we finally observe
\begin{align}
% \LL S + S \LL^* &= \lim_{t\rightarrow\infty} 
%  \int_0^t \left( \LL e^{\LL(s-t)} \BB \BB^* e^{-\LL^* (s-t)} + e^{\LL(s-t)} \BB \BB^* e^{-\LL^* (s-t)} \LL^*\right ) ds \\
%   &= \lim_{t\rightarrow\infty} 
%  \int_0^t \left(  e^{\LL(s-t)} \BB \BB^* e^{-\LL^* (s-t)} \right )' ds \nonumber \\
%  &= \lim_{t\rightarrow\infty}  e^{\LL(s-t)} \BB \BB^* e^{-\LL^* (s-t)}\;\;  \Big|_0^t \nonumber \\
%  &= \BB \BB^*.  \nonumber 
    \LL S + S \LL^* &= \lim_{t\rightarrow\infty} 
    \int_0^t \left( \LL e^{\LL(s-t)} \BB \BB^* e^{\LL^* (s-t)} 
    + e^{\LL(s-t)} \BB \BB^* e^{\LL^* (s-t)} \LL^*\right ) ds \\
    &= \lim_{t\rightarrow\infty} 
    \int_0^t \left(  e^{\LL(s-t)} \BB \BB^* e^{\LL^* (s-t)} \right )' ds \nonumber \\
    &= \lim_{t\rightarrow\infty} e^{\LL(s-t)} \BB \BB^* e^{\LL^* (s-t)}\; \Big|_0^t \nonumber \\
    &= \BB \BB^*,  \nonumber 
\end{align}
% \Add{where prime indicates differentiation w.r.t.\ $s$.}
where prime indicates differentiation with respect to $s$.
This result, sometimes referred to as the fluctuation-dissipation theorem, characterizes the long-term behavior of an OUP.
% In fluctuating hydrodynamic models, it is applicable to linear SPDE and to the semidiscrete system of SDEs obtained by discretizing in space.
In fluctuating hydrodynamics models, it is applicable to linear stochastic partial differential equations and to the semidiscrete system of stochastic ordinary differential equations obtained by discretizing in space. 
It can also be applied to the Fourier transformed equations to obtain the static structure factor.

\section{\texttt{StochasticHeat} Python program}

Available as a Jupyter Notebook from \url{github.com/AlejGarcia/IntroFHD}.

%%%%%%%%%%%%%%%%%%%%%%%%%%%%%%%%%%%%%%%%%%%%%%%%%%%%%%%%%%%%%%%
{\scriptsize

\begin{verbatim}


# StochasticHeat - Program to calculate trajectories for the stochastic heat equation 

# Set up configuration options and special features
import numpy as np                 # Use numerical python library
import matplotlib.pyplot as plt    # Use plotting python library
# Display plots in notebook window
%matplotlib inline                 

#* Set option flags and key parameters for the simulation
BC_FLAG =  1                  # Boundary condition flag (0: Periodic; 1: Dirichlet)
NONEQ_FLAG = 1     # Non-equilibrium flag (0: Thermodynamic equilibrium; 1: Temperature gradient)
if( BC_FLAG == 0 and NONEQ_FLAG == 1 ):
    print('ERROR: Periodic boundary conditions are only for thermodynamic equilibrium')
    NONEQ_FLAG = 0            # Reset the flag to equilibrium
PERTURB_FLAG = 1              # Start from perturbed initial condition (0: No; 1: Yes)
STOCH_FLAG = 1                # Thermal fluctuations? (0: No, deterministic; 1: Yes, stochastic)
SCHEME_FLAG = 1               # Numerical scheme (0: Forward Euler; 1: Predictor-corrector)

Ncells = 32                   # Number of cells (including boundary cells '0' and 'Ncells-1' )
MegaSteps = 8                 # Number of simulation timesteps in millions
# Recommend at least 2 million steps for equilibrium systems, 8 million for non-equilibrium systems

#* Set physical parameters for the system (iron bar)
kB = 1.38e-23       # Boltzmann constant (J/K)
mAtom = 9.27e-26    # Mass of iron atom (kg)
rho = 7870.         # Mass density of iron (kg/m^3)
c_V = 450.          # Specific heat capacity of iron (J/(kg K))
ThCond = 70.        # Thermal conductivity of iron (W/(m K))
Length = 2.0e-8     # System length (m)
Area = (2.0e-9)**2  # System cross-sectional area (m^2)

#* Set numerical parameters (time step, grid spacing, etc.).
dx = Length / Ncells       # Grid size (m)
dV = Area*dx               # Volume of grid cell (m^3)
    
xc = np.zeros(Ncells)
for ii in range(Ncells):
    xc[ii] = (ii +0.5) * dx       # Cell center positions (m)

kappa = ThCond / (rho * c_V )                # Coefficient in deterministic heat equation
# Coefficient in stochastic heat equation
if( STOCH_FLAG == 1 ):
    alpha = np.sqrt( 2 * kB * kappa / (rho * c_V) )   
else:
    alpha = 0.  # Set alpha = 0 to turn off thermal fluctuations

stabilityFactor = 0.1                         # Numerical stability if stabilityFactor < 1.
dt = stabilityFactor * dx**2 /( 2. * kappa )  # Timestep (s)

#* Set initial conditions for temperature
Tref = 300.      # Reference temperature (K)
if( NONEQ_FLAG == 0 ):
    T_Left = Tref; T_Right = Tref       
else:
    Tdiff = 400.   # Temperature difference across the system
    T_Left = Tref - Tdiff/2.; T_Right = Tref + Tdiff/2. 
    
# Standard deviation of temperature in a cell at the reference temperature
Tref_SD = np.sqrt(kB*Tref**2 / (rho * c_V * dV))

# Set initial temperature and its deterministic steady-state value
T = np.zeros(Ncells); T0 = np.zeros(Ncells)
for ii in range(0,Ncells):
    T0[ii] = T_Left + (T_Right-T_Left) * xc[ii]/Length      # Linear profile
    T[ii] = T0[ii] 

    if( PERTURB_FLAG == 1 ):
        T[ii] += Tref_SD*np.random.randn()  # Add random perturbation----

#* Summarize system parameters and initial state
print( 'System is iron bar with about ', int(rho*Length*Area/mAtom), ' atoms' )
print( 'System length = ',Length*1.0e9,' (nm); volume = ',
      (Length*Area)*1.0e27,' (nm**3)' )
print( 'Number of timesteps = ', MegaSteps,' million' )
print( 'Time step = ', dt*1.e15, ' (fs)' )
print( 'Number of cells = ', Ncells )
if( BC_FLAG == 0 ):
    print('** PERIODIC boundary conditions **')
else:
    print('++ DIRICHLET boundary conditions ++')

if( NONEQ_FLAG == 0 ):
    print('Equilibrium temperature = ',Tref,' Kelvin')
else:
    print('Fixed temperatures (K) = ',T_Left,' left; ',T_Right,' right')
print( 'At T = ',Tref,'K, standard deviation sqrt(< dT^2 >) = ', Tref_SD,'K' )
if( STOCH_FLAG == 0 ):
    print('******* DETERMINISTIC - NO THERMAL FLUCTUATIONS *****************')
if( SCHEME_FLAG == 0 ):
    print('-- Forward Euler Scheme --')
else:
    print('== Predictor-Corrector Scheme ==')

# Plot temperature versus position
plt.plot(xc, T,'g*',xc, T0,'b--')
plt.xlabel('Position (m)'); plt.ylabel('Temperature (K)')
plt.title('Initial state (stars), steady state (dashed)')
plt.show()

#* Initialize sampling plus misc. coefficients and arrays
Nsamp = 0                   # Count number of statistical samples
sumT  = np.zeros(Ncells)    # Running sum of T_i ; used to compute mean
sumT2 = np.zeros(Ncells)    # Running sum of (T_i)**2 ; used to compute variance
sumTT = np.zeros(Ncells)    # Running sum for correlation T_i * T_iCorr
sumSk = np.zeros(Ncells)    # Running sum for structure factor S(k)

iCorr = np.int_(Ncells/4)   # Grid point used for correlation

# Coefficients used in the main loop calculations
coeffDetFE = kappa * dt / dx**2
coeffStoFE = alpha * dt / dx
coeffZnoise = 1. / np.sqrt( dt * dV )

# Arrays used in the main loop calculations
Determ = np.zeros(Ncells); Stoch = np.zeros(Ncells+1)
Znoise = np.zeros(Ncells+1); Tface = np.zeros(Ncells+1)
PreT = np.copy(T)             # Used by Predictor-Corrector scheme
Spectrum = np.zeros(Ncells)   # Used to compute structure factor S(k)

#* Loop over the desired number of time steps.
NstepInner = 10                                       # Number of time steps (inner loop)
NstepOuter = np.int_(MegaSteps*1000000/NstepInner)    # Number of time steps (outer loop)
NskipOuter = NstepOuter/10        # Number of outer steps to skip before sampling begins
NdiagOuter = NstepOuter/20        # Number of outer steps between diagnostic outputs

for iOuter in range(NstepOuter):     # Outer loop

    # Print diagnostics
    if (iOuter % NdiagOuter) == 0 : 
        print( 'Finished ',np.int_(100*iOuter/NstepOuter),' percent of the time steps')
      
    for iInner in range(NstepInner):       # Inner loop

        # Deterministic update for temperature; first and last cells have BC unique stencils
        for ii in range(1,Ncells-1):
            Determ[ii] = coeffDetFE * ( T[ ii+1 ] + T[ ii-1 ] - 2*T[ ii ] )
        if( BC_FLAG == 0 ):
            Determ[0] = coeffDetFE * ( T[ 1 ] + T[Ncells-1 ] - 2*T[ 0 ] )  # Periodic BC
            Determ[Ncells-1] = coeffDetFE * ( T[ 0 ] + T[Ncells-2] - 2*T[Ncells-1] )  # Periodic BC            
        if( BC_FLAG == 1 ):
            Determ[0] = coeffDetFE * ( T[ 1 ] + 2 * T_Left - 3*T[ 0 ] )  # Left Dirichlet BC
            Determ[Ncells-1] = coeffDetFE * ( 2*T_Right + T[Ncells-2] - 3*T[Ncells-1] )  # Right BC
            
        # Generate random noise Z on cell faces; Znoise[ i ] is on left face of cell i
        Znoise = coeffZnoise * np.random.normal(size=Ncells+1)
       
        if( BC_FLAG == 0 ):
            Znoise[-1] = Znoise[0]     # Periodic BC
        if( BC_FLAG == 1 ):
            Znoise[0] = Znoise[0]*np.sqrt(2)    # Left Dirichlet BC
            Znoise[-1] = Znoise[-1]*np.sqrt(2)  # Right Dirichlet BC
            
        # Tface[ i ] is average between T of cells i-1 and i; value on the left face of cell i
        for ii in range(1,Ncells):
            Tface[ii] = 0.5 * (T[ ii-1 ] + T[ ii ])     
        if( BC_FLAG == 0 ):
            Tface[0] = 0.5*(T[Ncells-1] + T[0])   # Periodic BC
            Tface[Ncells] = Tface[0]              # Periodic BC
        else:
            Tface[0] = T_Left         # Dirichlet BC
            Tface[Ncells] = T_Right   # Dirichlet BC
        
        # Stochastic update for temperature
        for ii in range(Ncells):
            Stoch[ii] = coeffStoFE * ( Tface[ii+1]*Znoise[ii+1] - Tface[ii]*Znoise[ii] )

        if( SCHEME_FLAG == 0 ):
            # Forward Euler scheme
            for ii in range(Ncells):
                T[ii] += Determ[ii] + Stoch[ii]   # Total update for temperature
        else:
            # Predictor-Corrector scheme
            for ii in range(Ncells):
                PreT[ii] = T[ii] + Determ[ii] + Stoch[ii]     # Predictor step
            
            # Corrector step
            for ii in range(1,Ncells-1):
                Determ[ii] = coeffDetFE * ( PreT[ ii+1 ] + PreT[ ii-1 ] - 2*PreT[ ii ] )
            if( BC_FLAG == 0 ):
                Determ[0] = coeffDetFE * ( PreT[ 1 ] + PreT[Ncells-1 ] - 2*PreT[ 0 ] )  # Periodic
                Determ[Ncells-1] = coeffDetFE * ( PreT[ 0 ] + PreT[Ncells-2] - 2*PreT[Ncells-1] )  
            if( BC_FLAG == 1 ):
                Determ[0] = coeffDetFE * ( PreT[ 1 ] + 2 * T_Left - 3*PreT[ 0 ] )  # Dirichlet BC
                Determ[Ncells-1] = coeffDetFE * ( 2*T_Right + PreT[Ncells-2] - 3*PreT[Ncells-1] )
            
            for ii in range(1,Ncells):
                Tface[ii] = 0.5 * (PreT[ ii-1 ] + PreT[ ii ])    
            if( BC_FLAG == 0 ):
                Tface[0] = 0.5*(PreT[Ncells-1] + PreT[0])   # Periodic BC
                Tface[Ncells] = Tface[0]   # Periodic BC
            else:
                Tface[0] = T_Left
                Tface[Ncells] = T_Right
                
            for ii in range(Ncells):
                Stoch[ii] = coeffStoFE * ( Tface[ii+1]*Znoise[ii+1] - Tface[ii]*Znoise[ii] )
                
            for ii in range(Ncells):
                T[ii] = 0.5*( T[ii] + PreT[ii] + Determ[ii] + Stoch[ii] )              


    # End of Inner loop

  # Take statistical sample
    if( iOuter > NskipOuter ):
        Nsamp += 1
        for ii in range(Ncells):
            sumT[ii] += T[ii]            # Running sum for temperature average
            sumT2[ii] += T[ii]**2        # Running sum for temperature variance
            sumTT[ii] += T[ii]*T[iCorr]  # Running sum for temperature correlation
            
        if( NONEQ_FLAG == 0 ):
            # Take Fourier transform and record sampled spectrum in running sum
            Spectrum[0:Ncells] = np.abs( np.fft.fft( T[0:Ncells] ) )**2
            for ii in range(Ncells):
                sumSk[ii] += Spectrum[ii]

# End of Outer loop

#* Calculate average, variance, and correlation
aveT = np.zeros(Ncells)     # Average < T_i >
varT = np.zeros(Ncells)     # Variance < (T_i - <T_i>)**2 >
corrT = np.zeros(Ncells)    # Correlation < T_i T_iCorr >

for ii in range(Ncells):
    aveT[ii] = sumT[ii] / Nsamp
    varT[ii] = sumT2[ii]/Nsamp - aveT[ii]**2
for ii in range(Ncells):
    corrT[ii] = sumTT[ii]/Nsamp - aveT[ii]*aveT[iCorr]

#* Plot temperature versus x
plt.plot(xc, T,'g*',xc,aveT,'ro',xc,T0,'b--')
plt.xlabel('Position (m)'); plt.ylabel('Temperature (K)')
plt.title('Final state (stars), Average (circles), Deterministic (line)')
plt.show()

#* Plot variance of temperature versus x; compare with theory
varT_Th = np.zeros(Ncells)   # Theoretical value
if( NONEQ_FLAG == 0 ):
    varT_Th[0:Ncells] = kB*Tref**2 / (rho * c_V * dV)  
else:
    varT_Th[0:Ncells] = kB*T0[0:Ncells]**2 / (rho * c_V * dV)
    
# Conservation of energy correction for periodic boundary case
if( BC_FLAG == 0 ):
    varT_Th *= (1 - 1/(Ncells)) 
    
plt.plot(xc, varT,'ro',xc,varT_Th,'b--')
plt.xlabel('Position (m)'); plt.ylabel('Variance < dT**2 >')
plt.title('Average (circles), Equilibrium theory (dashed)')
plt.show()

#* Plot temperature cell correlation versus x
if( NONEQ_FLAG == 0 ):
    varTheory = kB*Tref**2 / (rho * c_V * dV)  # Temperature variance in cell iCorr
else:
    varTheory = kB*T0[iCorr]**2 / (rho * c_V * dV)  # Temperature variance in cell iCorr

if( BC_FLAG == 0 ):
    corrT_Th = - varTheory * 1/(Ncells) * np.ones(Ncells)  # Conservation of energy correction
    corrT_Th[iCorr] = varTheory * (1 - 1/(Ncells))         # for periodic boundary case
else:
    corrT_Th = np.zeros(Ncells)                              # Kronecker delta correlation
    corrT_Th[iCorr] = varTheory                              # for Dirichlet boundary case

if( NONEQ_FLAG == 1 ):
    for ii in range(Ncells):
        # Non-equilibrium correlation; see J. Stat. Phys. 47 209 (1987)
        corrNonEq = kB*(T_Right-T_Left)**2 / (rho * c_V * Area * Length**3)
        if( ii < iCorr ):
            corrNonEq *= xc[ii]*(Length - xc[iCorr])
        else:
            corrNonEq *= xc[iCorr]*(Length - xc[ii])   
        corrT_Th[ii] += corrNonEq

plt.plot(xc, corrT,'ro',xc,corrT_Th,'b--')
plt.xlabel('Position (m)'); plt.ylabel('Correlation < dT(x) dT(x*) >')
plt.title('Average (circles), Theory (dashed)')
plt.show()
print('Correlation for x* = ', xc[iCorr],' (m)')

#* Plot temperature cell correlation versus x WITHOUT the equilibrium delta function
corrTd = np.array(corrT)
corrTd[iCorr] = np.nan        # Delete data point at iCorr using NaN
corrTd_Th = np.array(corrT_Th)
corrTd_Th[iCorr] = np.nan        # Delete data point at iCorr using NaN                                                            

plt.plot(xc, corrTd,'ro',xc,corrTd_Th,'b--')
plt.xlabel('Position (m)'); plt.ylabel('< dT(x) dT(x*) > [delta-function removed]')
plt.title('Average (circles); Theory (dashed)')
plt.show()

#* Compute the measured structure factor and compare with linear theory
if( NONEQ_FLAG == 0 ):     # Only compute S(k) for equilibrium 
    NN = len(range(Ncells))
    Sk = np.zeros(NN)                           # Power spectrum S(k), AKA the structure factor
    Sk[0:NN] = sumSk[0:Ncells] / Nsamp     # Average S(k)
    Nyq = np.int_(np.floor(NN/2))               # Nyquist frequency is kSpect[NNNy+1]
    ki = np.arange(1,Nyq)                       # Wavenumber index, skipping the zero index
    Sk_eq = varTheory*NN * np.ones(Nyq+1)       # Equilibrium structure factor, which is constant
    
    # Plot structure factor, skipping the k=0 wavenumber index
    plt.plot( ki, Sk[1:Nyq],'ro', ki, Sk_eq[1:Nyq],'b--',)
    plt.xlabel('Wavenumber index'); plt.ylabel('Structure factor S(k)')
    plt.title('Average (circles); Theory (dashed)')
    plt.show()
else:
    print('Structure factor not computed when system is not at thermodynamic equilibrium')

\end{verbatim}
}
%%%%%%%%%%%%%%%%%%%%%%%%%%%%%%%%%%%%%%%%%%%%%%%%%%%%%%%%%%%%%%%

\bibliographystyle{siamplain}
\bibliography{MAIN_Bib}